\documentclass[11pt,oneside,letterpaper]{article}
\usepackage{amssymb}
\usepackage{amsmath}
\usepackage[dvips]{graphicx}
\usepackage{setspace}
\usepackage{amsfonts}
\usepackage{fancyhdr}
\usepackage{xcolor}
\usepackage{graphicx}
\usepackage{rotating}
\usepackage{comment}
\usepackage{color}
\usepackage{cite}
\usepackage{braket}

\usepackage{tikz}

\usepackage{subfigure}
\usetikzlibrary{decorations.pathmorphing}

\definecolor{darkgreen}{rgb}{0,0.5,0}
\definecolor{darkblue}{rgb}{0,0,0.6}
\definecolor{purple}{rgb}{0.4,.2,0.7}

\newcommand{\be}{\begin{equation}}
\newcommand{\ee}{\end{equation}}

\usepackage[colorlinks=true,citecolor=darkgreen,linkcolor=black,urlcolor=purple]{hyperref}

\usepackage{pdfsync}

\makeatletter
\newcommand*{\defeq}{\mathrel{\rlap{%
                     \raisebox{0.3ex}{$\m@th\cdot$}}%
                     \raisebox{-0.3ex}{$\m@th\cdot$}}%
                     =} 
\makeatother

\def\be{\begin{eqnarray}}
\def\ee{\end{eqnarray}}

\newcommand{\bea}{\begin{eqnarray}}
\newcommand{\eea}{\end{eqnarray}}
\def\ben{\begin{equation}}
\def\een{\end{equation}}

     \let\r=v

\def\be{\begin{equation}}
\def\ee{\end{equation}}
\def\ba{\begin{array}}
\def\ea{\end{array}}

\def\ba#1\ea{\begin{align}#1\end{align}}
\def\bs#1\es{\begin{split}#1\end{split}}

\allowdisplaybreaks  

\numberwithin{equation}{section}

\def \be {\begin{equation}}
\def \ee {\end{equation}}
	
\def \JM#1 {{\color{darkgreen}  JM: #1 }}
\def \SK#1 {{\color{red}  SK: #1 }}
\def \YC#1 {{\color{blue}  YC: #1 }}

\begin{document}
\onehalfspacing

\begin{center}

~
\vskip5mm

{\LARGE  {TF1 Snowmass Report: } } 
\\ ~ \\
{\LARGE{Quantum gravity, string theory, and black holes \\
}}

\vskip20mm

Daniel Harlow$^{1}$, 
Shamit Kachru$^{2}$,
Juan Maldacena$^{3}$, 
Ibrahima Bah$^{4}$,
Mike Blake$^{5}$,
Raphael Bousso$^{6}$,
Mirjam Cvetic$^{7}$,
Xi Dong$^{8}$,
Netta Engelhardt$^{1}$,
Tom Faulkner$^{9}$,
Raphael Flauger$^{10}$,
Dan Freed$^{11}$,
Victor Gorbenko$^{2}$,
Yingfei Gu$^{12}$,
Jim Halverson$^{13}$,
Tom Hartman$^{14}$,
Sean Hartnoll$^{15}$,
Andreas Karch$^{11}$,
Hong Liu$^{1}$,
Andy Lucas$^{16}$,
Emil Martinec$^{17}$,
Liam McAllister$^{14}$,
Greg Moore$^{18}$,
Nikita Nekrasov$^{19}$,
Sabrina Pasterski$^{20}$,
Monica Pate$^{21}$,
Ana-Maria Raclariu$^{20}$,
Krisha Rajagopal$^{1}$,
Shlomo Razamat$^{22}$,
Steve Shenker$^{2}$,
Sakura Schafer-Nameki$^{23}$,
Gary Shiu$^{24}$,
Eva Silverstein$^{2}$,
Douglas Stanford$^{2}$,
Brian Swingle$^{25}$,
Wati Taylor$^{1}$,
Nicholas Warner$^{26}$,
Beni Yoshida$^{20}$

\vskip15mm

{\it $^{1}$ MIT, Cambridge, Massachusetts, USA } \\
 {\it $^{2}$ Stanford University, Palo Alto, California, USA } \\
 {\it $^{3}$ Institute for Advanced Study, Princeton, New Jersey, USA } \\
 {\it $^{4}$ Johns Hopkins University, Baltimore, Maryland, USA}\\
{\it $^{5}$ University of Bristol, Bristol, United Kingdom}\\
{\it $^{6}$ University of California, Berkeley, California, USA}\\
{\it $^{7}$ University of Pennsylvania, Philadelphia, Pennsylvania, USA}\\
{\it $^{8}$ University of California, Santa Barbara, California, USA}\\
{\it $^{9}$ University of Illinois, Urbana-Champaign, Urbana, Illinois, USA }\\
{\it $^{10}$ University of California, San Diego, California, USA}\\
{\it $^{11}$ University of Texas at Austin, Austin, Texas, USA}\\
{\it $^{12}$ California Institute of Technology, Pasadena, California, USA}\\
{\it $^{13}$ Northeastern University, Boston, Massachusetts, USA}\\
{\it $^{14}$ Cornell University, Ithaca, New York, USA}\\
{\it $^{15}$ DAMTP, University of Cambridge, Cambridge, United Kingdom}\\
{\it $^{16}$ University of Colorado, Boulder, Colorado, USA}\\
{\it $^{17}$ University of Chicago, Chicago, Illinois, USA}\\
{\it $^{18}$ Rutgers University, Piscataway, New Jersey, USA}\\
{\it $^{19}$ SCGP, Stony Brook University, Stony Brook, New York, USA}\\
{\it $^{20}$ Perimeter Institute for Theoretical Physics, Waterloo, Ontario, Canada}\\
{\it $^{21}$ Harvard University, Cambridge, Massachusetts, USA}\\
{\it $^{22}$ Technion - Israel Institute of Technology (IIT), Haifa, Israel}\\
{\it $^{23}$ Mathematical Institute, University of Oxford, Oxford, United Kingdom}\\
{\it $^{24}$ University of Wisconsin, Madison, Wisconsin, USA}\\
{\it $^{25}$ University of Maryland, College Park, Maryland, USA}\\
{\it $^{26}$ University of Southern California, Los Angeles, California, USA}\\

\vskip5mm

\end{center}

\vspace{4mm}

\begin{abstract}
\noindent
 We give an overview of the field of quantum gravity, string theory and black holes summarizing various white papers in this subject that were submitted as part of the Snowmass process.


 
 \end{abstract}

\pagebreak
\pagestyle{plain}

\setcounter{tocdepth}{2}
{}
\vfill
\tableofcontents

\newpage


 \section{Quantum gravity and the nature of spacetime } 
 
 
 The twentieth century started with two revolutions in physics: quantum mechanics and relativity. 
 Quantum mechanics explains how matter works at the microscopic level, while relativity gives a new understanding of space and time. Over the century it was understood how to combine quantum mechanics with special relativity, resulting in the general framework of quantum field theory. This framework has been spectacularly successful at explaining the nature of matter, with the Higgs boson discovery being the latest landmark.  It has also given a general method for understanding many-body quantum phenomena such as superconductivity and the quantum Hall effect.  However, the search for a theory that can put together general relativity with quantum mechanics is still ongoing. We know how to perform a semiclassical quantization of general relativity coupled to field theory in the framework of effective field theories. This has been  a very useful tool for analyzing problems that involve relatively low curvatures, including physics in the vicinity of large black holes or during inflation in the early universe.  Unfortunately, however this theory is not renormalizable, which means that we need to introduce new unknown constants at each order in perturbation theory.  
 This makes it impossible to use this method to study situations where the spacetime curvature is not small in Planck units. The most important situation of this type is the initial singularity of the universe. Even if we include inflation or eternal inflation, our understanding would not be complete without describing this initial singularity. Similarly, there are singularities in the interiors of black holes, which again cannot be understood without a formulation of quantum gravity which is more robust than the semiclassical expansion.

In addition to the obvious problems of the cosmological and black hole singularities, there are also more subtle questions whose resolutions seem to require a theory of quantum gravity.  For example, the semiclassical quantization of gravity strongly suggests that a black hole should be viewed as a quantum system with an entropy given by its horizon area in Planck units, but the semiclassical description does not seem to be able to describe the degrees of freedom which give rise to this entropy.  This gives rise to Hawking's famous black hole information problem, which argues that we have to either give up the idea that this entropy really counts microstates in the usual way or else give up the validity of our semiclassical quantization even in regimes where the curvature is small.  Another problem along these lines arises in the context of inflation, which is currently our best candidate for a theory of the origin of structure in the universe.  Namely, most theories of inflation are ultimately theories of eternal inflation, which in the semiclassical approximation gives rise to a vast inhomogeneous universe where conventional notions of probability and prediction lead to bizarre paradoxes.  It thus seems that we need to either give up a successful theory of structure formation or else understand how these paradoxes are resolved in a complete theory of quantum cosmology. 

There is also a more 
logical reason to pursue a theory of quantum gravity: it is simply not acceptable to have two theories of physics, one for small things and one for big things, without knowing how the two theories fit together.  There are many precedents of this kind of situation in science, such as the separate theories of electricity and magnetism, or, more recently, the extension of quantum mechanics by the inclusion of special relativity.  The problem is theoretically hard, which shows that mere logical consistency is a very strong constraint. So far, no complete theory was found -- we do not have even one theory that fits all the data. Therefore,  one can say that we are limited by theory rather than by experiment; more theory is thus what is needed.  Nature has given us a number of hints: the existence of gravity and quantum mechanics, black holes, a positive cosmological constant, inflation, etc, and it is our job to find a theory that accommodates all of them.

A full theory of quantum gravity is likely to also have important implications for the rest of physics.  For example, it may shed light on some of the currently puzzling mysteries of particle physics, such as the reason for the gauge groups and the three families of matter particles, the values of the coupling constants, the nature of dark matter, the solution of the strong CP problem, etc. Perhaps not all of these problems have a quantum gravity explanation, but some of them might. In addition, we can hope to learn more about the physics of the early universe. If inflation happened, what sets the possible parameters of the potential?  Are alternatives to inflation possible? Is there a time before the singularity?

In addition to these possible direct applications, quantum gravity has already proved to be quite valuable as a source of new ideas about a wide range of seemingly-unrelated problems in physics and mathematics.  Ideas from black hole physics have led to many insights into the theory of strongly-interacting quantum many-body systems such as quark-gluon plasma and condensed matter systems, and the geometry of string compactifications and supersymmetric quantum field theories have opened up entire new branches of mathematics.  There has also been a rich interplay with the theory of quantum information and computation.  From this point of view quantum gravity, and in particular, string theory, has already become an indispensable part of the broader physics enterprise.   

There are a variety of approaches to quantum gravity, including string theory, loop quantum gravity, asymptotic safety, causal dynamical triangulations, etc.  For the most part, we will focus on string theory in this report. String theory has succeeded in reproducing four-dimensional general relativity at low-energies, for example, the Lorentz-invariant four-graviton scattering amplitude. It has also given a microscopic interpretation to the Bekenstein-Hawking formula for black hole entropy in some cases. However, the problem of quantizing gravity is difficult and multifaceted, and there is a broader set of ideas that will continue to influence the development of the subject.  Some of the results we discuss in later sections rely only on low-energy effective field theory, and thus should hopefully be valid in any self-consistent theory of quantum gravity.  Further discussion of some additional ideas and approaches appears in the white papers \cite{Giddings:2022jda,deBoer:2022zka,Berglund:2022kew,Berglund:2022qcc,Mottola:2022tcn,Mottola:2022rul}.   

In the remainder of this report, we will review recent progress on all of the above topics, and also discuss promising directions for upcoming research.  We will aim to incorporate perspectives from a number of community white papers on these topics, while also injecting some of our own perspectives.  By necessity, we will have to focus on some topics more than others, but our goal is to paint a broad picture of what is happening and where it might be going.

 \section{String theory and particle physics} 
 
String theory  naturally accommodates general features of the standard model of particle physics such as non-abelian gauge fields, chiral fermions, and multiple generations of matter fields. 

 The best understood limits of string theory involve closed (and sometimes also open) strings propagating
 in flat, supersymmetric ten-dimensional space-time (though versions also exist in other dimensions, and are certainly worthy of exploration).  They allow for manifestly ultra-violet finite computations of
gravitational scattering, curing one of the thorny problems of perturbative general relativity.  Unfortunately,  we do not find ourselves in maximally-supersymmetric flat space in ten dimensions, so for phenomenological reasons, it is useful to study versions of string theory where some of the
 dimensions are curled up on a compact manifold $X$.  
 Happily, it turns out that details of the topology and geometry of $X$ play an important role in determining the
 resulting lower-dimensional physics.  This provides the attractive possibility that open questions in 
 particle physics phenomenology or cosmology may be transformed into geometric questions about the extra
 dimensions.  Since structural questions about the Standard Model abound -- including the possibility of gauge coupling unification,  the nature of dark matter, the explanation of striking hierarchies in the Yukawa couplings of matter fields, the absence of CP violation in the strong interactions, and the origin of neutrino masses -- one can hope to find a new layer of explanations in the higher-dimensional theory. We elaborate on research in this direction here.

Efforts to connect string theory to elementary particle physics have a rich history, reviewed in \cite{Cvetic:2022fnv} (where one can find many relevant references).
A great deal of the effort in research to date has been focused on making tree-level supersymmetric solutions of string theory that realize N=1 supersymmetric models of
(close to) realistic particle physics.  Supersymmetry breaking is left as a further refinement.
Restricting first to this general setting, one can investigate a suite of complementary questions:
\begin{itemize}
\item[1)] Is it possible to realize particle physics broadly similar to that of the Standard Model, for instance with the gauge group of the Minimal Supersymmetric Standard Model and the right spectrum of chiral gauge representations?  
 
\item[2)] Is it possible to refine this by finding models which exhibit desired structures of e.g. Yukawa couplings, or which have gauge coupling unification?

\item[3)] Some but not all questions in phenomenology are necessarily sensitive to ultra-violet physics.  
 Beyond the hierarchy problem, which here is being solved by supersymmetry, these include questions about inflation and dark energy.  Some models of supersymmetry breaking and its mediation to the Standard Model fields also have UV sensitivity. To what extent can string theory accomodate suggested models of these UV sensitive phenomena?  Does it introduce new ideas? Does it significantly constrain the space of
 ideas?
 \end{itemize}
 On question 1), there has been progress in waves. 
 Of the supersymmetric string theories, the heterotic theories and the type I theory already enjoy large gauge groups in ten dimensions.  In the mid 1980s, it was realized that
 these theories can give rise to MSSM-like spectra via compactification on Calabi-Yau manifolds.  Unification (with Wilson-line breaking of the GUT group) is very natural in this framework.
 
 The type II theories were viewed as phenomenologically unpromising in the 1980s, as they lack obvious sources of non-Abelian gauge groups and charged matter.  These problems are both cured by the realization that the theories contain D-branes, and inclusion of either these branes or suitable geometric singularities can yield realistic gauge groups. As a result, crudely realistic models became viable on e.g. intersecting stacks of D-branes by the late 1990s.  
 
 The more modern era has seen a refined understanding of the non-perturbative ``F-theory" description of type IIB string theory
 give rise to a plethora of MSSM-like constructions via engineering appropriate singularities in Calabi-Yau geometries. (Roughly, geometric singularities in F-theory capture dynamics of coincident 7-branes from the
 perturbative IIB string theory view.)  In fact, the classification of promising geometries for Standard Model - like physics becomes -- in this setting -- a very concrete (but challenging) problem in algebraic geometry. Unification may be possible in this framework.
 
 Very large numbers of candidate models exist in the various construction methods described (as well as in M-theory compactifications on spaces of $G_2$ holonomy, where the picture is still being developed).  One promising direction of active research is studying large classes of these models at once with the goal of understanding what sorts of particle physics properties (i.e. gauge groups, number of generations, etc) are typical and atypical.  It may prove natural to use new numerical techniques, e.g. machine learning, to address such questions.  
 
 On question 2), there has been technical progress in translating the computation of Yukawa couplings to (sometimes independently interesting) questions in algebraic or enumerative geometry. The central physics question is to explain the patterns of small and large couplings; because Yukawa couplings are marginal, the explanation can either lie in very high energy physics or in physics at lower scales. In the earliest heterotic models, Yukawa couplings are related to intersection numbers of homology cycles within the extra dimensions, and topological selection rules could be used to explain large dimensionless ratios of Yukawas.   More recently, in type II models, attempts to explain Yukawa hierarchies have  helped inspire the exploration of novel types of instanton effects in string theory.  
 
 Grand unification seems very natural in the heterotic picture (where the MSSM gauge group naturally arises as a subgroup of the larger ten-dimensional gauge group).  It is less natural in intersecting brane models, and may be achievable in F-theory models with 7-branes.  In all cases, details of making the unification work are subtle.  (Current data is also not conclusive on the question of whether achieving unification should be a central goal or not.)

 On question 3), where supersymmetry breaking plays a crucial role, there has been a large volume of work but there remains room for substantial progress.  The issue immediately becomes entangled with that of understanding the potential for the ``moduli fields" -- the scalar fields which govern the size and shape of the extra dimensions, the locations of branes, etc.  Supersymmetry breaking must involve generation of a potential energy function, which will generically depend on these fields and lead to possible instabilities as well as interesting opportunities for cosmology.

 At a technical level, a systematic understanding -- even at the level of the close-to-supersymmetric systems we are currently discussing -- will necessarily involve precise understanding of superpotential computations in string theory. 
 Some superpotential terms, arising from fluxes in the extra dimensions, are readily computed using classical techniques from mathematics (e.g. computation of period integrals on Calabi-Yau spaces).  Others, involving non-perturbative effects, are still harder to understand in detail.
 Models of inflation and dark
 energy are discussed at greater length in \S3 (on quantum gravity and cosmology) here, and in the white paper \cite{Flauger:2022hie}.  They require a level of theoretical control of effective potentials which remains just at the forefront of present technology. A variety of interesting scenarios and ideas have been and are being explored, and there is active debate about what is and isn't reasonable to expect in the particle physics and cosmology of string compactifications. One area of recent progress is the numerical computation of Calabi Yau metrics and other geometrical data of the relevant manifolds, which may eventually give a concrete handle on the low-energy Kahler potential, which has important applications in both particle phenomenology and cosmology.    
 
 So far, we have focused on the discussion of supersymmetric models of particle physics, presuming supersymmetry to be broken at low energies.  There is no evidence that supersymmetry is realized in Nature, and it is very well motivated to study string models with supersymmetry broken at higher energy scales (with natural possibilities being the Kaluza-Klein scale, where one sees extra dimensions, or the string scale).  This introduces new challenges (as issues of stability of scalar fields present in the tree level low-energy spectrum come to the fore immediately).  As such challenges need to be faced after supersymmetry breaking in any case, it seems that this is a less explored direction where future developments promise to be very interesting. Fruitful such investigations have already been occurring in the interface of string theory with cosmology.
 
 One important result of these explorations is that a relatively large number of axion like fields seems fairly common. These axions could have a range of masses. One of them could be involved in explaining the smallness of the QCD $\theta$-angle, but others could also be present. This spurred experiments to looks for such non-QCD axions in a wide range of masses and couplings, ranging from table top to using rotating black holes as axion detectors. 
 
 A particular particle physics scenario (alternative to low-energy supersymmetry) that has been a special focus of interest is the one introduced by Randall and Sundrum (see the white paper \cite{Agrawal:2022rqd}).  
 In this scenario, a slice of a highly warped 5d space-time is used to explain the hierarchy; the Standard Model Higgs fields can naturally live in the region of small warp factor.  Such warped geometries are common in examples of AdS/CFT, and a dual description of their scenario involves a (cut-off) conformal field theory coupled to gravity.  
 This scenario may enjoy a natural home in string theory: fluxes can give rise to significant warping, and for this reason large hierarchies of scales are easily accessible in warped regions of string compactifications. There has been a large body of research on this, and should data indicate it to be a leading scenario in particle physics going forward, one can anticipate continued development.


 \section{String theory and cosmology}

 In this section we summarize the conclusions of the white paper \cite{Flauger:2022hie},  where an extensive list of references can be found. 
 
 Quantum gravity interfaces cosmology in a number of ways. One is that it could constrain the type of effective field theories that we can consider, both now as well as in the early universe.
 The expansion of the universe stretches space in such a drastic way that microscopic distances in the early universe are expanded to cosmological scales. For this reason the large scale structure of our universe can carry    information about physics at extremely short scales or high energies, energies higher than those available to particle colliders.   
 
In string theory, the value of the dark energy depends on the shape of the internal dimensions, as well as possible branes and/or magnetic-like fluxes that are present there.
The constructions that are simplest to analyze are those that have zero dark energy and have unbroken supersymmetry. 
However, there has been vigorous ongoing research to construct more realistic configurations with positive values of the dark energy and broken supersymmetry,  with one of the first plausible scenarios being \cite{Kachru:2003aw}. Several objections and counter objections to such scenarios have been raised, and the effort to make them more concrete is still ongoing. This involves the intricate mathematics of Calabi Yau manifolds, as well as sophisticated control of non-perturbative corrections.  Other compactifications developed in parallel yield power law stabilization mechanisms, including more general curved manifolds which are also amenable to concrete mathematical analysis.  These constructions indicate that string theory has a vast landscape of possible solutions, each of them with distinct four dimensional physics.  There has been a significant effort at understanding the constraints on the effective four dimensional theories that come from such constructions, see \cite{Harlow:2022gzl} for a recent review of some of these ideas.  
As mentioned in the previous section, one rather generic feature of these constructions is the appearance of a possibly large number of axion-like fields. In addition to their relevance for particle physics, such particles are also useful for constructing inflationary models and may contribute a time-dependence to dark energy in the present universe.

  One  question of great experimental and theoretical interest is whether quantum gravity can produce inflationary models with relatively large values of $r$, the tensor to scalar ratio. Future experiments promise to greatly tighten the upper bound on $r$ down to around $r \leq 10^{-3}$ \cite{CMB-S4:2016ple,LiteBIRD:2022cnt}. So the question is whether quantum gravity, or string theory, allow models with $r> 10^{-3}$, which are values potentially measurable in the near future. 
  The value of $r$ is related to the total amount of displacement experienced by the inflaton field as it slowly rolls down the potential \cite{Lyth:1996im}. 
 At the level of effective field theories we can have relatively large values of $r$, and the simple power law potentials have already been excluded by the present constraints. 
 In  string theory, the first models suggested a very small value for $r$, but subsequently other proposals suggested larger values, see \cite{Baumann:2014nda,Silverstein:2016ggb}. However, it is possible that there is a theoretical upper bound for $r$ and it would be very interesting to understand where it is.   
 
 Some of the microscopic models for inflation contain extra light fields, of masses of order the Hubble scale during inflation. 
 Conventional cosmological perturbation theory can be used to study the imprints that these extra particles could have on the primordial fluctuations. It turns out that they give rise to non-gaussian features. 
 Already the presence of gravity leads to a very small non-gaussian signal in the simplest single field model. For this reason,  cosmological correlation functions have been computed in perturbation theory. This problem involves essentially a type of scattering amplitude in de Sitter, which is intimately connected to its cousins in   both flat space and Anti-de Sitter.
 
 Some specific microscopic inflationary models lead to other potentially observable features, such as oscillations in the potential, or interesting non-gaussian tails in the primordial fluctuations, or a reduced speed of sound during inflation,   \cite{Baumann:2014nda,Silverstein:2016ggb}.

 Turning to more conceptual questions,  an interesting puzzle is how to deal with eternal inflation.   In particular, the ``measure problem'':  how should we compute probabilities in eternal inflation? 
 A closely related question is whether there is a wavefunction of the universe, and if so how it might be computed.   This is an area which seems to need a deeper input from quantum gravity, since there are proposals that vary widely. The most mathematically elegant is the Hartle-Hawking proposal, which unfortunately seems to be in contradiction with experiment.  Along these lines there have been attempts to extend lessons from AdS/CFT to de Sitter space, but we are lacking a sufficiently explicit holographic example which leads to a standard Einstein-like gravity in the bulk. It seems likely that some of the recent progress in the description of quantum black holes \cite{Bousso:2022ntt} could have implications for cosmology, and we now turn to this subject.

 \section{Quantum aspects of black holes and the emergence of spacetime} 
 One of the most important aspects of quantum gravity is the quantum mechanics of black holes.  Semiclassical calculations strongly suggest that black holes should have entropy 
\be\label{SBH}
S=\frac{\mathrm{Area}}{4G},
\ee
and also that a black hole of mass $M$ should evaporate over a timescale of order $t_{evap}\sim G^2 M^3$.
These results lead directly to Hawking's information problem: any information about how the black hole was made lies deep in the interior of the black hole, which the semiclassical picture tells us is spacelike-separated from the near-horizon region where the radiation is being produced, and thus the evaporation process destroys this information in contradiction with the unitarity of quantum mechanics.  It has sometimes been hoped that unitarity could be restored by ``small corrections'' to Hawking's calculation, but by now it is clear that no small modification of semiclassical evolution can restore unitarity \cite{Mathur:2009hf}.  We thus need to either accept that the evaporation process is not unitary, which would itself lead to a violation of \eqref{SBH}, or else that the semiclassical picture of spacetime is badly violated by the microscopic description of black holes.  Said differently, the spacetime near black holes, and thus spacetime everywhere, may well be \textit{emergent}: it is not a fundamental part of the theory, and only exists in some situations and to some approximation.  This topic has been the subject of a large amount of recent work, which was reviewed extensively in the white paper \cite{Bousso:2022ntt} and whose main points we now summarize (see \cite{Bousso:2022ntt} for many more references).

In string theory there are various limits where a non-perturbative description of quantum gravity can be obtained. These limits have been used to give precise countings of black hole microstates, which have matched expectations from black hole thermodynamics to ever-increasing precision (see  \cite{Dabholkar:2014ema} for a particularly impressive recent match).    The best-understood of these limits is the AdS/CFT correspondence, which says that quantum gravity in asymptotically-AdS space is equivalent to conformal field theory living at its asymptotic boundary.  Initial studies of this correspondence were focused on using classical gravity in the ``bulk'' AdS description to study strongly-coupled dynamics in the boundary theory (see the following section), but in the last 15 years much progress has been made on understanding in more detail how the bulk spacetime emerges from the boundary description.  A key early breakthrough was the Ryu-Takayanagi proposal, eventually refined into the quantum extremal surface (QES) formula, for computing the von Neumann entropy of a boundary state in terms of bulk quantities.  This led to a broader understanding that boundary entanglement is a key element of the emergence of the bulk, which was eventually expressed mathematically in the idea that the correspondence should be understood as a quantum error-correcting code \cite{Harlow:2018fse}. In the last few years this program led to the remarkable discovery that the ``Page curve'' of an evaporating black hole can be reliably computed using the QES formula, with the result being just what is expected from a unitary evaporation process \cite{Almheiri:2020cfm}. 

In parallel a deep connection was developed between the dynamics of black holes and quantum chaos.  It was realized that simple gravitational calculations can serve as probes of chaos in the boundary description, in particular giving a reliable method for computing the ``scrambling time'' of black holes using out-of-time-order correlators.  Moreover it was understood that subtle features in the analytic continuation of the thermal partition to complex temperature, usually packaged into ``spectral form factor'' $|Z(\beta+it)|^2$, can be understood either as consequences of random matrix theory in the boundary description  or from subleading saddles in the gravitational path integral.  It was realized that many of these ideas can be explored quite explicitly in the context of the SYK model, giving a way to test the chaotic nature of black holes in a solvable regime \cite{Saad:2018bqo}.   

An additional interesting direction has pursued a relationship between computational complexity and the emergence of spacetime.  The basic idea is that bulk operations which are exponentially complex (in the black hole entropy) do not need to respect the semiclassical structure of spacetime, while those of sub-exponential complexity do need to respect it. This idea is still not in its final form, but it has been explored in many different avatars over the last decade, leading to a number of novel insights, see eg.  \cite{Brown:2019rox}. 

Despite all this recent progress, there are still many open questions about the emergence of spacetime near black holes.  One such question is to what extent the interior remains smooth for black holes which are sufficiently ``typical''.  Some time ago it was observed in \cite{Braunstein:2009my,Mathur:2009hf,Almheiri:2012rt,Almheiri:2013hfa,Marolf:2013dba} that there are serious obstacles to realizing a conventional description of the black hole interior in a microscopic theory where the entropy of the black hole is given by \eqref{SBH}.  Some of these obstacles have since been overcome, but the general situation is still unknown and we seem to still be missing some key ingredient.  Relatedly we do not yet know how to think about the black hole singularity in a precise way: is there any sharp question we can ask about it, and in particular is there any sense in which sufficiently small objects which fall into it pass through and end up somewhere else? More generally we still have the problem that our best techniques for analyzing black holes in the gravitational variables are still only approximate, so there are boundary observables (such as the black hole S-matrix or the detailed late-time behavior of the spectral form factor) which we are not able to account for using known bulk techniques.  One idea is to study such questions using string theory, but string theory remains in most cases only a perturbative framework so doing this would require a deeper formulation of string theory than we currently have.

This leads us naturally to another white paper \cite{Bena:2022ldq}, which reviews the ``fuzzball'' program  attempting to explicitly construct the black hole microstates as fully back-reacted, gravitational solutions of string theory.  Large families of solutions have been found, in some cases with a number  approaching  the one given by the area formula. Extending these results to Schwarzschild black holes remains an important challenge.   
 
We are optimistic that the quantum mechanics of black holes will continue to be a source of progress and inspiration in the coming decade, bringing together insights from many different branches of theoretical physics.

 \section{New ideas for  many-body systems from quantum gravity and black holes}

 In this section we summarize the white paper \cite{Blake:2022uyo}. 
 The AdS/CFT correspondence relates strongly coupled quantum systems to gravity. Thermal systems are related to spacetimes with black holes, or black branes. 
 This can be used to learn about strongly interacting   systems, since one can perform computations relatively easily in  the gravity description. This has led to some interesting general lessons, including  bounds on transport \cite{Policastro:2001yc}, a bound on chaos \cite{Maldacena:2015waa},   hydrodynamics with anomalies \cite{Erdmenger:2008rm,Banerjee:2008th,Son:2009tf}, etc. In some of these cases, the gravity analysis inspired  conjectures which were later proven in general,  without relying on gravity. The gravity theory served as a useful ``laboratory'' to make the more general discovery.
 
 An important application of these ideas is to the study of the quark gluon plasma. This deconfined phase of QCD is qualitatively similar to ${\cal N}=4 $ supersymmetric gauge theory at finite temperature. While the latter is not the same as QCD, the advantage of being able to do reliable computations at strong coupling has led to interesting insights. This is particularly true for real time, out of equilibrium problems, such as transport properties, jet energy loss, etc. 
 
 The gravity description has been very useful for studying far out of equilibrium phenomena, such as the evolution of a quantum system after a rapid change in parameters (a quench),  or the evolution of a strong localized perturbation. One surprising conclusion has been that the hydrodynamic description can become applicable after relatively short times of order the inverse temperature of the final state.

 Another important motivation was the study of strange metals, the materials that lead to high temperature superconductivity. It has been proposed that this system is related to nearly extremal black holes, or black branes,  with an $AdS_2$ near horizon geometry, which gives a new perspective on this long-standing problem.  The study of  
 this connection has also given back  interesting lessons for black holes. In particular the SYK model mentioned in the previous section \cite{Sachdev:1992fk,kitaevfirsttalk}, which was originally motivated by condensed matter physics, has been very useful for identifying certain symmetries that govern the gravitational dynamics of near extremal black holes, leading to a precise quantization of the most important low energy modes, as well as their non-perturbative structure as discussed in the previous section.   

 The study of quantum chaos is a well-developed subject, but the connection to black holes mentioned in the previous section led to a new diagnostic, the out-of-time-order correlator, which can be applied to study chaos in any many-body quantum system  \cite{Shenker:2013pqa,kitaevfirsttalk}.  This has led to a number of novel insights.  One of these is the idea that there is a characteristic velocity, the ``butterfly velocity'', at which chaos propagates through a system.  Another is a general bound $\lambda \leq 2\pi   T   $ on the growth of chaos, where $\lambda$ is a quantum version of a Lyapunov exponent which appears in the out-of-time-order correlator.  This bound is saturated by black holes, and seems to hold for rather general quantum systems \cite{Maldacena:2015waa}.  This bound is similar to the ``Planckian'' bound on many-body transport $1/\tau \lesssim T  $
 \cite{Zaanen:2018edk},
 which appears to be realized in the linear in $T$ resistivity of high temperature superconductors.


 Central to many of these developments are ideas of entanglement, tensor networks and quantum information. We will not discuss them here since they will be discussed in more detail by TF10 (Theory Frontier 10: quantum information), see in particular the white paper \cite{Faulkner:2022mlp}.
 
\section{String theory and mathematics}

 Historically, new theories of gravity have evolved hand-in-hand with important mathematical developments.  Newton (and Leibniz) famously developed calculus as the proper understanding of classical mechanics evolved in the 17th century, and Einstein's theory builds heavily on foundations of curved-space geometry laid by Gauss, Bolyai, Lobachevsky, Riemann and others.  One may therefore expect the proper understanding of quantum gravity to come hand in hand with novel mathematical developments.  This has proven to be the case in string theory.  In this section, we focus on the interplay of modern understanding of string compactification with developments in mathematics.  The rough idea will be the following.  Since string theory is a theory of gravity, string compactification
can be viewed as providing a rich set of probes of its behavior: one can use compactifications on various geometries $X_i$, and their relationships to one another, to probe how this particular theory of quantum gravity ``sees" spatial geometry and topology.  This viewpoint has led to the discovery of many new mathematical phenomena and structures (including mirror symmetry, vertex algebras, moonshine \cite{Harrison:2022zee}, aspects of the geometry of the moduli space of Riemann surfaces associated with string perturbation theory \cite{Berkovits:2022ivl}, and too many others to list here).
We will have to mention just a few out of many promising developments, described more completely in \cite{Bah:2022xfv} and \cite{Bah:2022wot}.
 
 Calabi-Yau compactification, and its generalization to compactifications on manifolds of special holonomy useful in M-theory compactification, motivates many mathematical questions.  The realization that Calabi-Yau manifolds come in mirror pairs (which are topologically distinct, but give rise to identical physics in a highly non-trivial way) originated in string theory, and led to striking developments in enumerative geometry.  This program has culminated in the intense investigation of homological mirror symmetry, which may lead to direct construction of mirror pairs by mathematically rigorous techniques.  
 
 The systematic computation of protected terms in the low-energy effective action of compactified string theories with various amounts of supersymmetry continues to lead to new mathematical developments.
 Study of cases with 4d $\mathcal{N}=2$ supersymmetry and computation of the related prepotentials led to exciting developments in Gromov-Witten theory (counting holomorphic spheres embedded in the compact dimensions as well as higher genus analogues).
 Looking forward, in a compactification on $X$ preserving 4d $\mathcal{N}=1$ supersymmetry, the computation of space-time superpotentials involves new questions in enumerative geometry of $X$.  (As examples,
 the superpotential can be related to counts of associative submanifolds in manifolds of $G_2$ holonomy, or divisors of arithmetic genus one in Calabi-Yau fourfolds.)  These lead to natural target generalizations of the questions that have been so fruitfully explored in the past.  
 
 String theory has also been a source of observations and conjectures about relations of natural objects in number theory to geometry.  This is related to the discussion above.
 For instance,  in simple cases, the supersymmetry-protected couplings in the effective action arising in string
 compactification on $X$ are both counting functions for geometric objects embedded in $X$, and enjoy special automorphic properties related to the geometry of the moduli space of Ricci-flat metrics on $X$.  This brings in interesting connections to the theory of automorphic forms.  Similarly, in string theory one can make supersymmetric charged black holes by starting with suitable configurations of D-branes wrapping cycles in a Calabi-Yau space (and then continuing the configuration to strong coupling).  
 By counting the microscopic D-brane configurations, one can obtain an estimate of the black hole entropy.
 The counting functions whose Fourier coefficients capture the black hole entropy (as a function of charges) thereby obtain interpretations in enumerative geometry.  In simple cases they are once again -- sometimes for mysterious reasons -- interesting automorphic forms.
 
 The interplay of string theory with quantum field theory also leads to concrete mathematical questions of interest.  Via ``geometric engineering," one can realize a wide varieties of $d \leq 6$ supersymmetric field theories arising as singular geometries in string compactification on manifolds of reduced holonomy.  Classification of (supersymmetric) field theories is an extremely difficult problem to get a handle on in its own terms.  Classification of local singularities of Calabi-Yau or $G_2$ geometries is an a priori different question that allows one to bring in powerful new tools.  
 It is not clear precisely how classifying ``geometrically engineered" field theories is related to a full classification effort.  Nevertheless, it gives one a new handle on a very large class of theories.  Furthermore, the geometric picture allows in many cases for a solution of the low-energy field theory, by turning questions about the low-energy effective action into (soluble) counting problems in the singular geometry.
 
 Another way of studying field theories via an associated geometry is provided by holography.  The AdS/CFT correspondence turns a wide class of conformal field theories -- often difficult to study by direct field theory techniques -- into AdS (super)gravity solutions.  Questions about the moduli space of marginal deformations of the field theory (or a larger space accessible also by relevant deformations) turn into questions about solutions of partial differential equations capturing the dynamics of bulk fields in the gravitational dual.  This leads to a possibility of new fruitful connections between the study of certain classes of non-linear PDEs and quantum field theory.
 
 Calabi-Yau spaces play a role as toy models of relevance in many questions about string phenomenology and physical mathematics.  As Yau's theorem is non-constructive, it is an interesting question in differential geometry to determine the form of the Ricci flat metric on a compact Calabi-Yau.  There is progress on this question, both by using numerical methods and advances in machine learning, and in the the special case of hyperK\"ahler metrics (relevant for e.g. K3 surfaces) by analytical methods as well.  Continued progress can be expected.
 
 Entirely different mathematical issues can arise in thinking about string theories where supersymmetry is broken at higher energies.  For instance, properties of strings on hyperbolic manifolds have been a subject of several interesting investigations.  This is a promising area for future research.

\section{New ideas for classical gravity from quantum gravity}
The study of quantum gravity has also spurred progress in classical gravity.  One example of this is the discovery, using AdS/CFT, of the ``fluid-gravity correspondence'', which shows that in certain regimes Einstein's equations reduce the Navier-Stokes equations of hydrodynamics \cite{Hubeny:2011hd}.  Many new solutions of Einstein's equations have also been found in the context of AdS/CFT, including black holes with scalar hair  and stationary black holes with non-Killing horizons \cite{Dias:2015nua}.  There has been progress understanding possible astrophysical signatures of near-extremal Kerr black holes based on conformal symmetry \cite{Gralla:2016jfc,Gralla:2016qfw}.  There has also been progress on the cosmic censorship conjecture, in both the negative direction, showing that in the absence of the weak gravity conjecture Einstein Maxwell theory in $AdS_4$ can violate the conjecture \cite{Horowitz:2016ezu,Crisford:2017gsb}, and in the positive direction, showing that the Penrose inequality, which is a well-known stand-in for cosmic censorship, can be derived in AdS/CFT \cite{Engelhardt:2019btp}.  An interesting connection between the Bondi, van der Burg, Metzner, and Sachs group of asymptotic symmetries in flat space, soft theorems, and the memory effect has also been discovered \cite{Strominger:2017zoo}, and attempts to develop this into a theory of holography in flat space were summarized in the white paper \cite{Pasterski:2021raf}.  This involves an interesting reformulation of the S-matrix where the Lorentz group is realized as the conformal group of the sphere at infinity.  The study of asymptotic symmetries has also motivated progress on the Hamiltonian formulation of gravity, in particular in the covariant phase space approach.  Substantial progress has also been made studying the properties of diffeomorphism-invariant observables in classical gravity, see also the white paper \cite{deBoer:2022zka}.  The close connection between work on classical and quantum gravity is likely to continue in coming years, with useful ideas flowing in both directions.


\vspace{1cm}
\textbf{Acknowledgments}

D.H. is supported by the Simons Foundation as a member of the ``It from Qubit'' collaboration, the Sloan Foundation as a Sloan Fellow, the Packard Foundation as a Packard Fellow, the Air Force Office of Scientific Research under the award number FA9550-19-1-0360, and the US Department of Energy under grants DE-SC0012567 and DE-SC0020360. S.K. is supported by the NSF under grant PHY-2014215; by the Simons Collaboration for Ultra Quantum Matter, which is a grant (651440, SK) from the Simons Foundation; and by a Simons Investigator Award.
J.M. is supported in part by U.S. Department of Energy grant DE-SC0009988, the Simons Foundation grant 385600.

\small
\bibliographystyle{ourbst}
 \bibliography{SnowmassTF1paper}
\end{document}